\documentclass[vecphys]{svmult}

\usepackage{makeidx}         
\usepackage{graphicx}        
\usepackage{multicol}        
\usepackage[bottom]{footmisc}

\makeindex             

\begin{document}

\title*{Barred Galaxies: an Observer's Perspective}
\author{Dimitri A. Gadotti}
\institute{Max Planck Institute for Astrophysics\\
Karl-Schwarzschild-Str. 1, D-85741 Garching bei M\"unchen, Germany\\
\texttt{dimitri@mpa-garching.mpg.de}}
%
%
\maketitle

\begin{abstract}
I review both well established and more recent findings on the
properties of bars, and their host galaxies, stemming from
photometric and spectroscopic observations, and discuss how these findings can be understood in terms of
a global picture of the formation and evolution of bars, keeping a connection with theoretical
developments. In particular, I show the results of a detailed structural analysis of $\approx300$
barred galaxies in the Sloan Digital Sky Survey, providing physical quantities, such as bar length,
ellipticity and boxyness, and bar-to-total luminosity ratio,
that can either be used as a solid basis on which realistic
models can be built, or be compared against more fundamental theoretical results. I also
show correlations that indicate that bars grow longer, thinner and stronger with dynamical age, and
that the growth of bars and bulges is connected. Finally, I
briefly discuss open questions and possible directions for future research.
\end{abstract}

\section{Some Basic Facts}
\label{sec:basic}

It is well established now that bars are very often found in disc galaxies. One usually finds a bar fraction of $\sim2/3$
\cite{esk00}, considering both prominent and weak bars, unless one is looking for bars in images observed
at too short wavelengths \cite{she03}, as the stellar content
of bars is usually dominated by old, red stars. In fact, most bars fade away in ultra-violet images \cite{paz06},
although some bars can still be recognised (cf. NGC 1097).
Even though the remaining $\sim1/3$ of disc galaxies do not seem to harbour identifiable bars, they might still have
less prominent non-axisymmetric distortions.

In addition, bars are seen in galaxies with a wide range of bulge-to-total ratio and mass,
i.e. from lenticulars to irregulars.
Thus, secular evolution processes induced by bars occur not only
in disc-dominated galaxies with inconspicuous bulges. They also happen in bulge-dominated galaxies, which
suggests the coexistence of classical, merger-built bulges, with bulges built from disc dynamical
instabilities \cite{kor04}. But secular evolution
can also happen without bars. Oval distortions in discs can also induce a substantial exchange of angular
momentum from the inner to the outer parts of galaxies \cite{kor04}.

Finally, a recent development suggests that the total amount of mass within stars that reside in bars at
$z\approx0$ is similar to that kept in stars belonging to
classical bulges. A similar amount is confined to elliptical galaxies.
Approximately $15\%$ of the total mass in stars at $z\approx0$ is located in bars.
Classical bulges and elliptical galaxies contain each a comparable fraction. This means that, as far as the stellar
mass budget in the local universe is concerned, bars are as relevant as classical bulges and elliptical galaxies.
The other $\sim1/2$ of the stellar mass content at $z\approx0$ is in galaxy discs \cite{gad08,dri07}.

\section{Observed Properties of Barred Galaxies}
\label{sec:galprops}

A number of theoretical studies indicate that bars redistribute the disc gas content,
driving it through torques and dynamical resonances into spiral arms,
rings and dust lanes, where it usually changes phase from neutral to molecular as it is submitted to higher pressures.
Gas lying beyond the bar ends is driven outwards, whereas gas lying within the bar ends is
driven to the central regions (e.g.
\cite{sch81,com85,ath92,fri93,pin95}; see also \cite{sel03,kor04}). This rearrangement
of the gas can erase chemical abundance radial gradients \cite{fri94}. Accordingly, barred galaxies
have flatter O/H gradients than unbarred galaxies \cite{zar94}, and the stronger the bar the flatter the gradient
\cite{mar94}. Furthermore, it has also been observed that the central concentration of molecular gas in barred galaxies
is higher than in unbarred galaxies \cite{sak99}.

What happens to the gas funnelled by the bar to the centre? Several observational results point out that star formation
activity is enhanced in the central regions of barred galaxies as compared to unbarred galaxies, for spirals of early
and intermediate types \cite{ser65,ser67,hua96,car97,ho97,alo01,jog05}, arguing
that the gas driven to the centre by the bar is relatively
efficiently transformed in young stars. Consequently, barred galaxies also
show flatter {\em colour} gradients than unbarred galaxies, due to bluer central regions \cite{gad01}. The average
{\em global} star formation, however, seems to be comparable in barred and unbarred galaxies
of intermediate types, since their integrated colours
are very similar. In addition, bars in late type spirals show star forming regions all along the bar, whereas bars
in early type spirals have star formation concentrated at the centre and/or at the bar ends \cite{phi96}.
There are suggestions that the former bars are also dynamically younger (i.e. formed later) than the latter
(\cite{fri95,mar97}; see also \cite{woz07,per07,ver07}).

Theoretical work indicates that not only the disc gas content is rearranged
by the effects induced by a bar, but that
the distribution of {\em stars} in the disc also changes (see e.g. \cite{pat03,pat06,rom07}). Furthermore,
due to an exchange of angular momentum between the disc and the dark matter halo, mediated by the bar,
the disc becomes more centrally concentrated \cite{ath02,ath03}.
The inflow of disc material (gas and stars) to the centre, and the star formation episodes associated with it,
seem to often produce sub-structures such as nuclear rings, nuclear discs and spiral arms, and secondary
bars, as shown by observations (see e.g. \cite{erw02}). The increase
in central mass concentration can also make stars migrate to orbits out of the disc plane \cite{ber07}.
These structures built at the centre of the disc by the bar
might naturally generate an excess of luminosity in the central part of the disc luminosity profile, as compared
to an extrapolation to the centre of its outer part.
Likewise, classical bulges can also be defined as the extra light on top of the inner disc profile,
and also extend out of the disc plane.
Thus, to distinguish these different components, formed through distinct
processes and having dissimilar physical characteristics, structures built through the inflow of disc
material are called pseudo-bulges \cite{kor04}, or disc-like bulges \cite{ath05}. There are also
observational evidences that bars might affect the distribution of stars in the outer disc \cite{erw07}.

Our understanding of yet another type of bulge also benefits largely from a successful connection between
theory and observation. Theoretical studies show that the vertical structure of a bar grows in time through
dynamical processes, generating a central structure out of the plane of the disc that can have a boxy or
peanut-shaped morphology (e.g. \cite{com81,rah91,mar06}). Several observational evidences, mostly based on
theoretical expectations, argue that such box/peanut bulges are indeed just the inner part of bars seen edge-on
(see \cite{des87,kui95,bur99,chu04,bur05}).

The gas brought to the centre by the bar can also end up fuelling a super massive black hole and AGN activity.
The basic idea is that the primary bar brings gas from scales of $\sim5$ kpc to $\sim1$ kpc and a secondary
bar instability brings gas further inwards to $\sim100$ pc. The gas still needs to loose further angular momentum,
and at that scales other physical processes, such as viscosity, come to help \cite{shl89,shl90}. However, studies
comparing the fraction of barred galaxies in quiescent and active galaxies show contradictory results
\cite{ho97,mul97,kna00,lai02,mai03}. Such comparisons are plagued by issues such as AGN classification, sample
selection and bar definition, which in turn depends on wavelength and spatial resolution. Should one expect
to see a clear distinction in bar fraction in quiescent and active galaxies? Even though there are more
clear evidences of bars at least building up a fuel reservoir at the galaxy centre for star formation and AGN activity
(\cite{sak99,mai99,dav04}; see also \cite{wys04}), the answer to this question is, for several reasons, more likely, no.
To begin with, we saw that bars are very often seen in disc galaxies, and thus any difference is likely to have low
statistical significance, unless samples are large enough. More important are factors such as the availability of gas
(quiescent barred galaxies might just lack gas to fuel the black hole) and the strength of the bar (or its ability to
bring gas inwards). The issue gets more complicated when one considers in detail the role of inner spiral arms and rings,
secondary bars and dynamical resonances near the centre. Inner rings might prevent gas to reach the
centre, and inner spiral arms could either remove or give further angular momentum from (to) the gas if they are
trailing (leading) \cite{com01}. Hydrodynamical simulations show that, at least in some cases, secondary bars might not help
(and might even prevent) the gas inflow to the nucleus (\cite{mac02}; but see \cite{hel07}). Finally, one must keep
in mind that the typical time-scale of AGN activity episodes is likely to be much shorter than the typical time-scale
for funnelling the gas and the life time of bars. This means that even if a bar is bringing gas to the nucleus
one has to be looking at the right time to see the black hole accreting the gas. Thus, one should not be too
surprised (or rather confused) to see results like the similar fractions of secondary bars found in quiescent and active
galaxies \cite{mar01}, or the fact that galaxies with the strongest bars are mostly quiescent \cite{shl00,lau02}.
Things are just not that simple! The nuclear region of barred galaxies shows complex structures that follow
several distinct patterns, which might be related to the strength of the bar and the final destiny of the
gas collected by it \cite{pee06}.

From theoretical work, it is still unclear whether primary and secondary bars are long-lived or have short lives.
Earlier studies suggested that primary bars could be destroyed by a central mass concentration, such as a super
massive black hole. However, it turned out that the central densities needed are unrealistically high, and that at most
a weakening of the bar can result (see \cite{she04,ath05+} and references therein). This
suggests that primary bars are
long-lived. Models that include accretion of gas from the halo to the disc,
however, suggest that primary bars can be destroyed
due to a transfer of angular momentum from the gas to the bar \cite{bou02,bou05}, but this is contested (see
\cite{ber07}). In addition, a new bar could be formed after the previous one vanished, since the accreted gas
could turn the disc unstable again, by replenishing the disc with new stars.
A key point in models that predict the demise and rebirth of bars is the availability of external gas.
They are thus just relevant to gas-rich galaxies. Still on theoretical ground, also the life of secondary bars
remains uncertain, with results suggesting both long and short life times (see \cite{mas97,elz03,deb06}).

A natural and direct way to assess the life time of bars is to study bars at different redshifts, but this is
as yet also not free from opposing results. Early studies had pointed out a sharp drop in the fraction of barred
galaxies at $z\sim1$ (see \cite{vdb96}). Not long afterwards, some studies showed that the apparent lack of bars was
only caused by band shifting and poor spatial resolution \cite{vdb02,she03,jog04,elm04}. The most recent studies,
however, do not seem to agree on whether there is a rapid decline in the bar fraction with redshift to $z\sim1$
\cite{she07}, or if this fraction is fairly constant \cite{bar07}. A constant fraction of barred galaxies from
$z\sim1$ to $z\sim0$ might put models that predict bar dissolution and reformation in trouble, as this would require
a fine tuning of the corresponding time-scales. There are recent indications that bars in early and intermediate type
spirals are long-lived \cite{elm07}. If this is correct, it indicates that, in fact, models of bar
reformation might concern only late type spirals (later
than Sc). This agrees with the findings in \cite{gad05,gad06}, where a methodology to estimate the dynamical ages
of bars is introduced. This will be discussed in more detail in the next section, including implications
for secular evolution scenarios and AGN fuelling by bars.

\section{Estimating the Dynamical Ages of Bars}
\label{sec:ages}

From the previous discussion it is possible to see that there is a number of questions whose answers are still
escaping from us. For instance:

\begin{itemize}

\item{How fast bars drive gas to the galaxy centre?}

\item{For how long are bars modifying the overall evolution of their host galaxies?}

\item{What is the fraction of stars in bulges that comes from secular processes?}

\item{When did the first bars appear?}

\item{Are bars recurrent? If yes, in which conditions?}

\end{itemize}

It is evident that a method through which one could estimate the dynamical ages of bars would greatly improve
tentative answers to these questions. One possibility, explored in \cite{gad05}, is to use the vertical
thickening of bars as a clock. As pointed out above, this is predicted by theoretical work and it is in
very good agreement with observations. The vertical extent of bars translates directly into the
vertical velocity dispersion of its stars, $\sigma_z$. Thus, one can take spectra of face-on barred
galaxies from the bar and from the disc and determine and compare the corresponding values of $\sigma_z$.
Recently formed bars should have values of $\sigma_z$ similar to that of the disc, from which it just formed.
Evolved bars should have $\sigma_z$ substantially higher than the disc. Two parameters have thus been defined
in \cite{gad05}:

\begin{itemize}

\item{$\sigma_{z,{\rm bar}}$ -- the vertical velocity dispersion of stars in the bar at a galactocentric
distance of about half the bar length; and}

\item{$\varDelta\sigma_z$ -- the difference between $\sigma_{z,{\rm bar}}$ and the vertical velocity dispersion
of {\em disc} stars at the same galactocentric distance.}

\end{itemize}

\noindent Using spectra obtained for a sample of 14 galaxies, and considering only clear cases, the authors find that
young bars have typically $\sigma_{z,{\rm bar}}\sim30$ km s$^{-1}$ and $\varDelta\sigma_z\sim5$ km s$^{-1}$, whereas
evolved bars have typically $\sigma_{z,{\rm bar}}\sim100$ km s$^{-1}$ and $\varDelta\sigma_z\sim40$ km s$^{-1}$.
Statistical tests indicate that these young and evolved bars are indeed different populations at 98\% confidence level.

Furthermore, with measurements of the length $L_{\rm Bar}$ and the colour $(B-I)_{\rm Bar}$ of these bars, presented
in \cite{gad06}, it is found that young bars have, on average, $L_{\rm Bar}\approx5.4$ kpc and
$(B-I)_{\rm Bar}\approx1.5$, whereas evolved bars have, on average, $L_{\rm Bar}\approx7.5$ kpc and
$(B-I)_{\rm Bar}\approx2.2$. Evolved bars are both longer and redder than young bars. The difference in length
also holds when it is normalised by the galaxy diameter. The bar colour was measured close to the bar
ends, but {\em outside} star forming regions.

The fact that evolved bars are longer than young, recently formed bars is in agreement with theoretical
results \cite{ath02,ath03,leo06}. These works indicate that, while bars evolve, they capture stars from the inner disc,
redistribute angular momentum along the disc and dark matter halo, and get longer and thinner in the process.
NGC 4608 and NGC 5701 might represent cases where the capture of disc stars by the bar is substantial.
Recent results suggest that the bar in NGC 4608 had an increase in mass of a factor of $\approx1.7$, through
the capture of $\approx13\%$ of disc stars (\cite{gad08}; see also \cite{gad03,lau05}).

The difference in colour between young and evolved bars represents a difference in the age of their stars of
$\approx10$ Gyr. As seen in the previous section, bars seem to follow two different patterns of star formation,
which might be related to the dynamical age. A recently formed bar seems to form stars along its whole extent,
whereas an evolved bar seems to form stars mostly at its centre and/or its ends.
This indicates that, when one carefully
measures the colour of stars in the middle of the bar, one is probing mainly
the first generation of stars formed in the
bar. That seems to be the reason why those bars which are {\em dynamically} old, as estimated from their stellar
kinematics, are also redder than the dynamically young bars. Alltogether, these results also
indicate that at least some bars are very old, and thus most likely not recurrent (unless the first generation
of bars has very short life times).

Interestingly, dynamically young bars are found preferentially in gas-rich, late type spirals. This suggests that
bar recurrence is restricted to this class of galaxies, as expected from such models (see discussion in previous
section). In addition, it was also found that galaxies hosting AGN typically have young bars, which possibly
means that the funnelling of gas to feed the black hole at the nucleus occurs on short time-scales.
A similar conclusion is reached in \cite{oht07} after the finding of a significantly higher bar fraction in
narrow-line Seyfert 1 galaxies, which are supposedly in an early stage of black hole evolution.

These results come, however, from the analysis of a small sample of galaxies. It is highly desirable to have
the dynamical ages of bars measured for a much larger sample, and assess the validity of these results.
Furthermore, at the current stage, one can only discriminate between recently formed and evolved bars.
It is now difficult to measure with more precision the dynamical age of the bar. Finally, one would like
to be able to give more stringent numbers to this parameter without having to rely on the age of the
bar stellar population. The accuracy in estimates of the dynamical age of a bar has to be improved. This might
be accomplished by an approach involving both observations (e.g. with a large scale 2D mapping of $\sigma_z$
in barred galaxies) and theory (e.g. with a more detailed analysis of the vertical evolution of bars
with time).

\section{The Structural Properties of Bars}
\label{sec:barstruct}

Although bars are ubiquitous and might account for a significant fraction of a galaxy total luminosity, only recently
studies dedicated to a more detailed modelling of the structural properties of bars started to come out more
often (see e.g. \cite{lau05,ree07,gad08}). This is partially because of the significant increase in complexity
when one includes another component in the structural modelling of disc galaxies. These studies usually make use
of the full data contained in 2D galaxy images, rather than only 1D surface brightness profiles, in order to obtain
more accurate results. The rise in complexity thus usually means that automated procedures become much less reliable.
In \cite{gad08}, {\sc budda} v2.1 \cite{des04} is used to individually fit galaxy images with model images that
include up to three components: a S\'ersic bulge, an exponential disc and a bar. Bars are modelled as a set of concentric
generalised ellipses \cite{ath90}, with same position angle and ellipticity:

\begin{equation}
\label{eq1}
\left(\frac{|x|}{a}\right)^c+\left(\frac{|y|}{b}\right)^c=1,
\end{equation}

\noindent where $x$ and $y$ are the pixel coordinates of the ellipse points, $a$ and $b$ are the extent of its
semi-major and semi-minor axes, respectively, and $c$ is a shape parameter. Bars are better described by boxy ellipses
(i.e. with $c>2$). The surface brightness profile of the model bar is described as a S\'ersic profile, as bulges.
The S\'ersic index of bars $n_{\rm Bar}$ is often in the range $\approx0.5-1$, with lower values representing
flatter profiles. Another bar parameter fitted by the code is the length of the bar semi-major axis $L_{\rm Bar}$,
after which the bar light profile is simply truncated and drops to zero.

\begin{figure}
\centering
\includegraphics[height=4cm,clip=true]{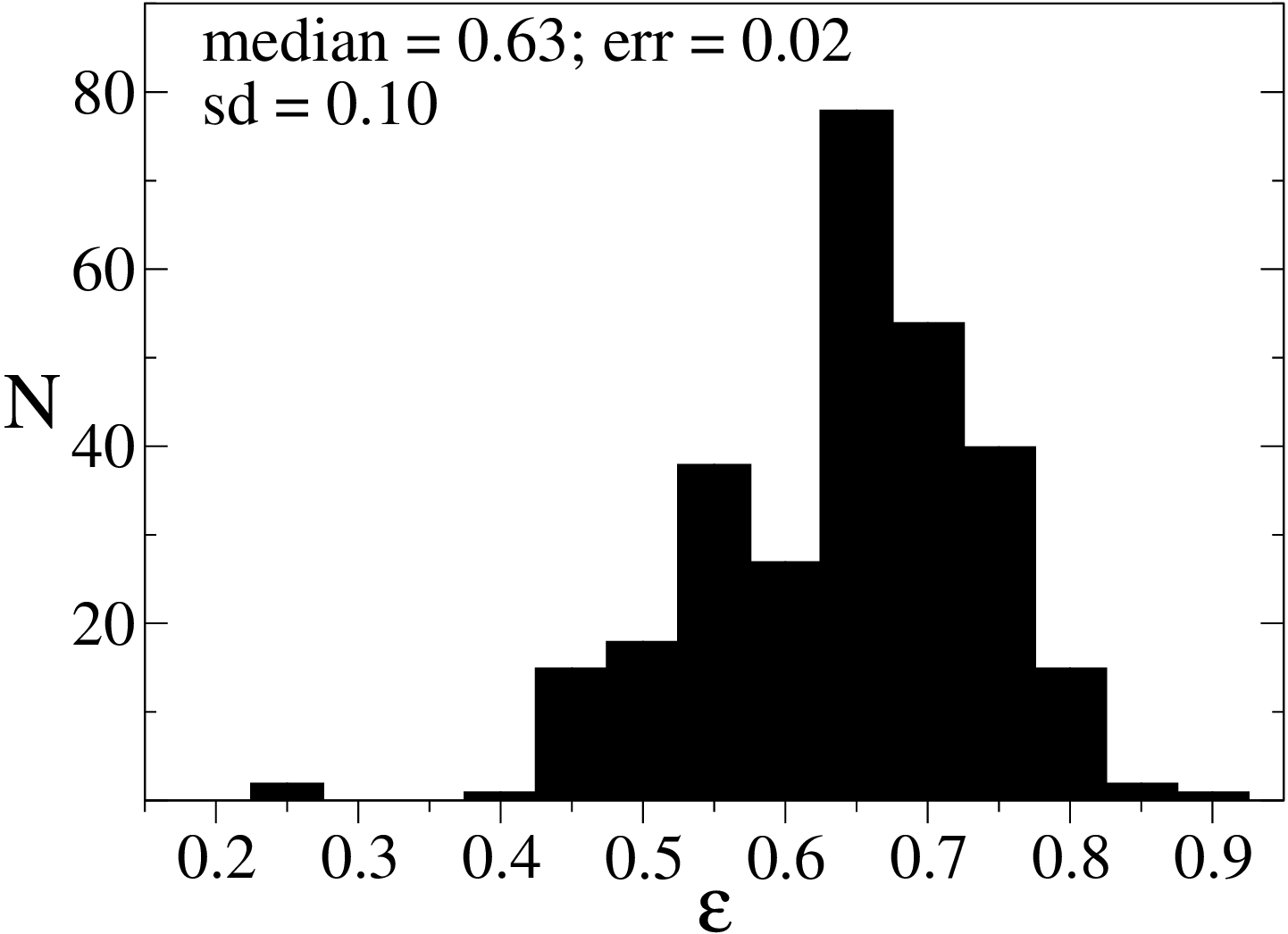}\hskip 0.2cm\includegraphics[height=4.1cm,clip=true]{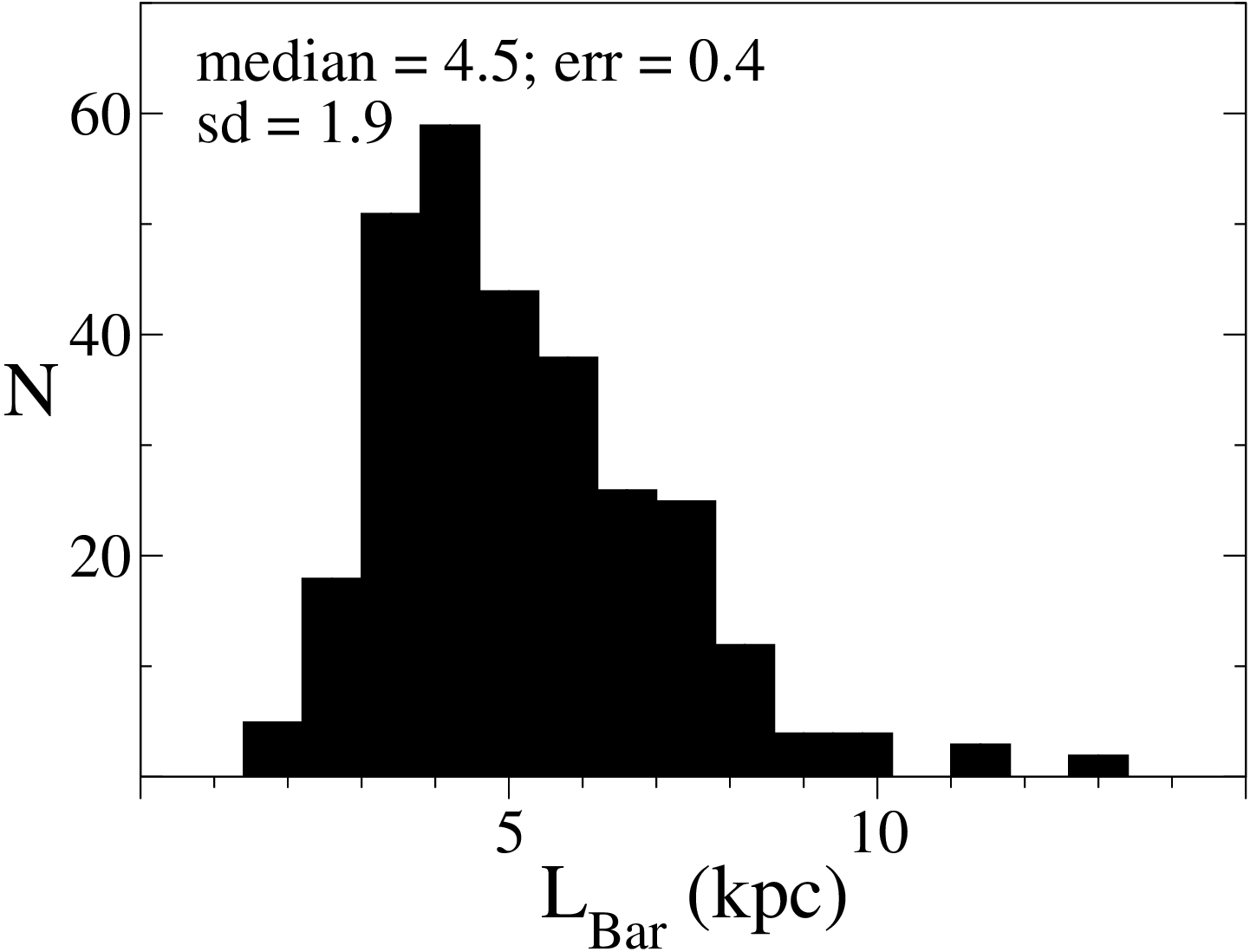}\\
\includegraphics[height=4.1cm,clip=true]{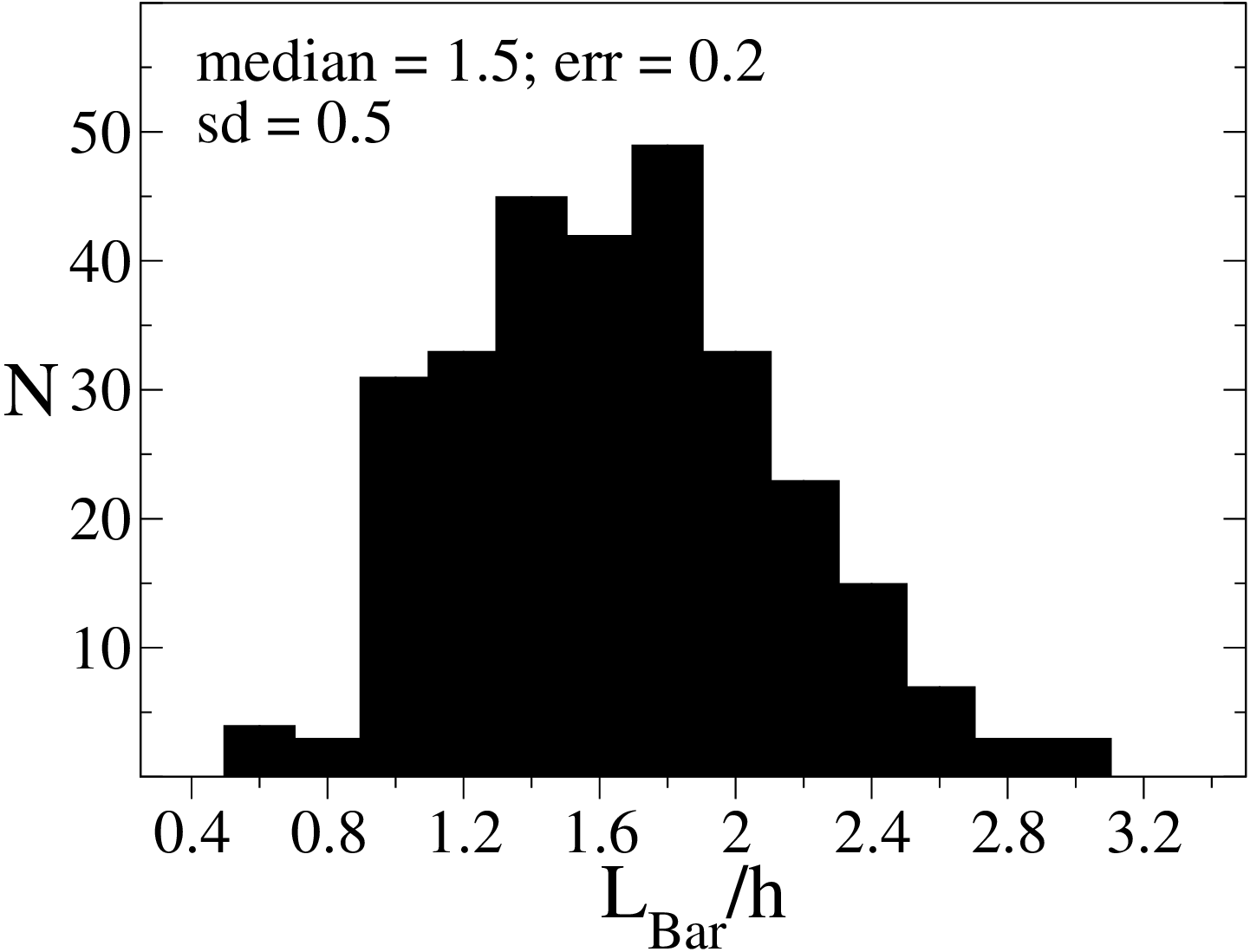}\hskip 0.2cm\includegraphics[height=4.1cm,clip=true]{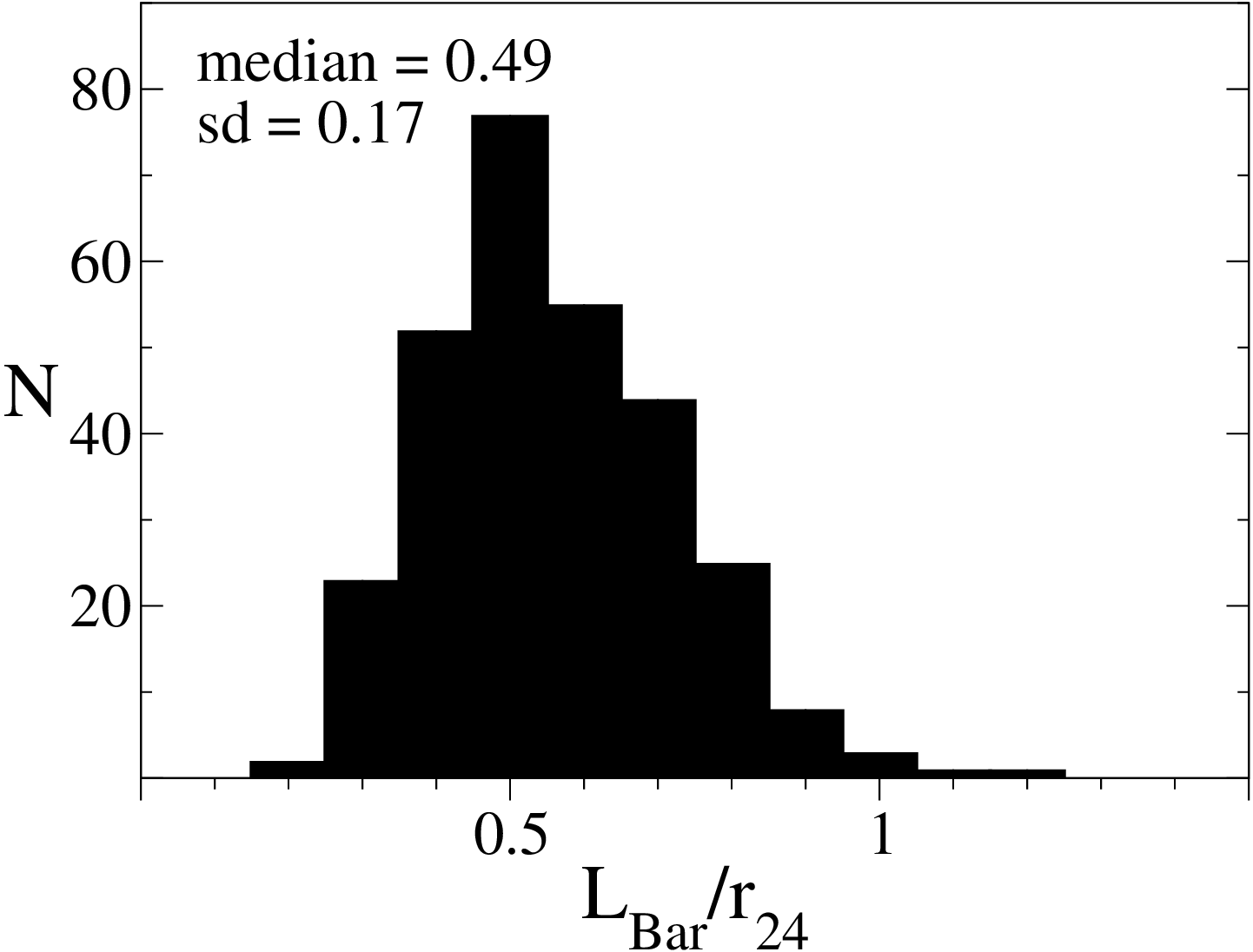}\\
\caption{Distributions of bar structural parameters obtained from 2D bar/bulge/disk $i$-band image decomposition
of $\approx300$ barred galaxies in the Sloan Digital Sky Survey (SDSS).
The sample spans from lenticular to bulgeless galaxies at
a typical redshift of 0.05, and excludes galaxies with stellar masses below $10^{10}$ M$_\odot$.
From top to bottom and left to right: bar ellipticity, bar length (semi-major axis in kpc
-- $H_0=75$ km s$^{-1}$ Mpc$^{-1}$), bar length normalised
by disc scalelength $h$, and bar length normalised by $r_{24}$ (the radius at which the galaxy surface brightness reaches
24 mag arcsec$^{-2}$ in the $r$-band). Noted at each panel are the median and standard deviation values of the
corresponding distribution, as well as the mean 1$\sigma$ error of a single measurement, when available.
Bin sizes are $\approx1-2\sigma$.}
\label{fig1}
\end{figure}

\begin{figure}
\centering
\includegraphics[height=4.1cm,clip=true]{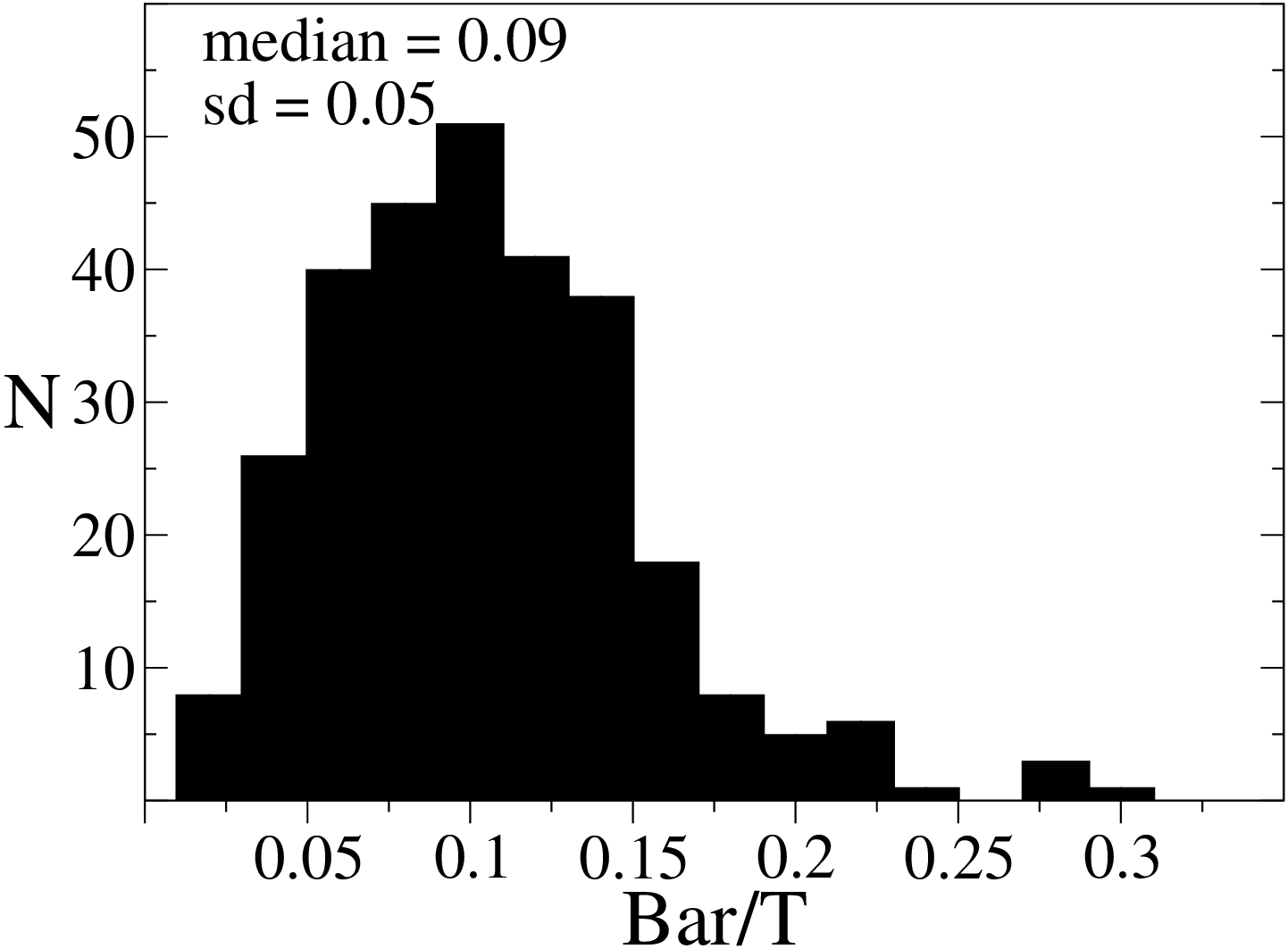}\hskip 0.2cm\includegraphics[height=4.1cm,clip=true]{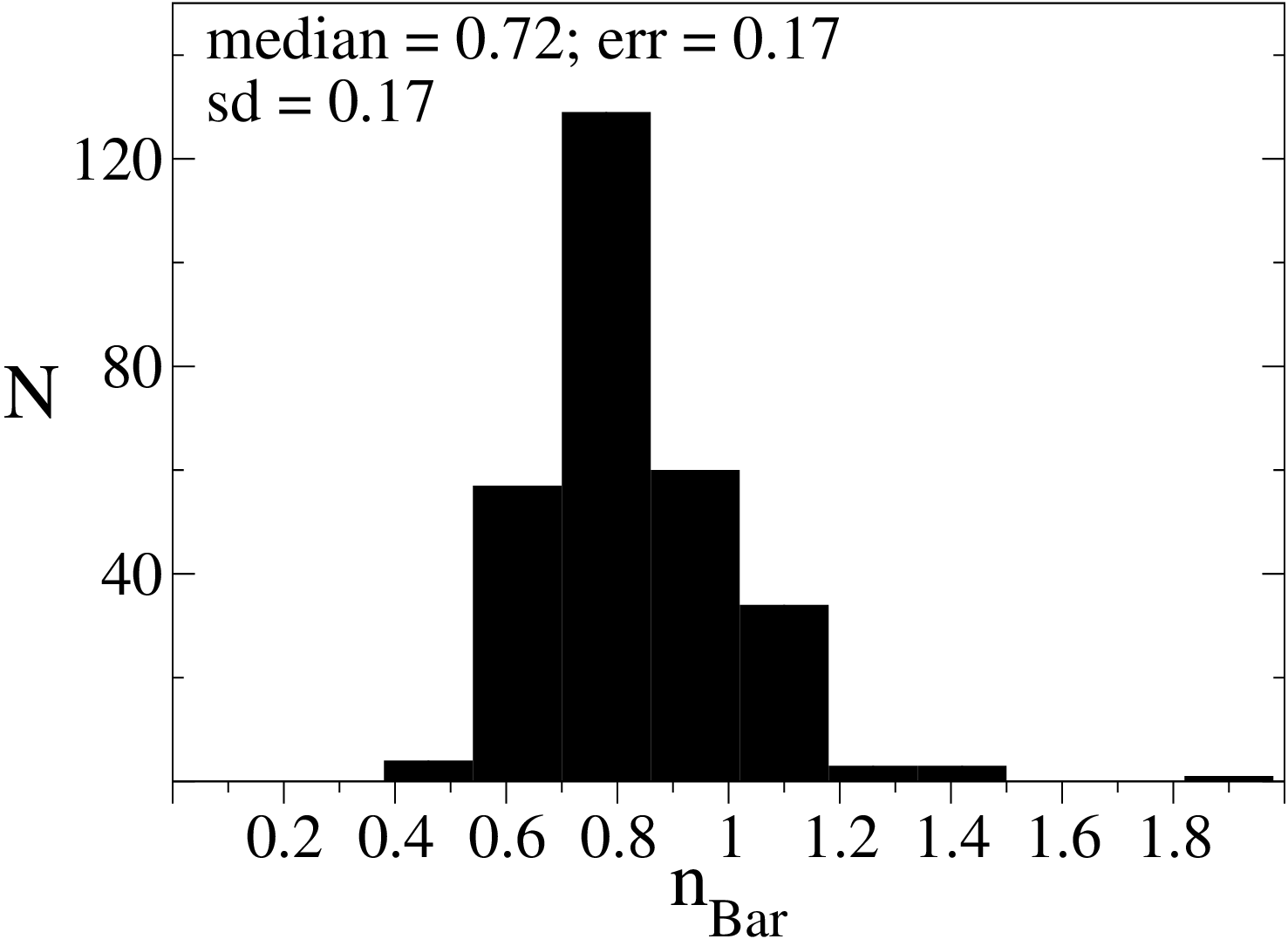}\\
\includegraphics[height=4.1cm,clip=true]{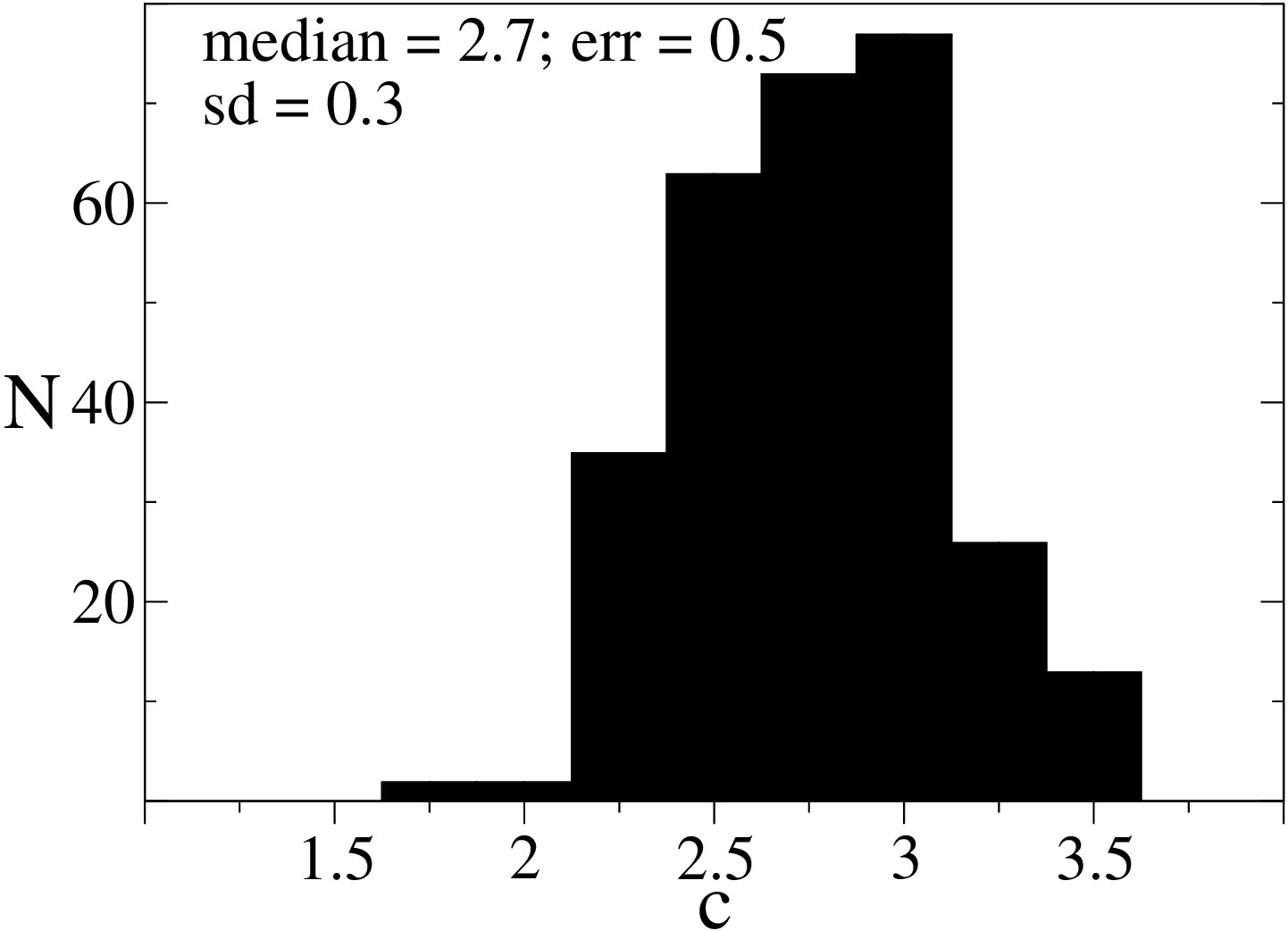}
\caption{Same as Fig. \ref{fig1}, but for bar-to-total luminosity fraction, bar S\'ersic index, and bar boxyness
[parameterised as $c$ -- see Eq. (\ref{eq1})].}
\label{fig2}
\end{figure}

Similar fits were individually done to $\approx1000$ galaxies in a sample carefully drawn from the
Sloan Digital Sky Survey (SDSS -- see \cite{gadlp}).
The sample spans from elliptical to bulgeless galaxies, with stellar masses
above $10^{10}$ M$_\odot$ (k-corrected $z$-band absolute magnitudes $\leq-20$ AB mag),
at a typical redshift of 0.05, and includes $\approx300$ barred galaxies. All galaxies
are very close to face-on (axial ratio $\geq0.9$)
and do not show morphological perturbations, thus assuring that dust extinction is minimised and
that the sample is suitable for image decomposition. This also avoids the uncertainties in obtaining deprojected
quantities. Fits were done in $g$, $r$ and $i$-band images.
The distributions of several bar structural parameters obtained
in this work from the $i$-band images are shown in Figs. \ref{fig1} and \ref{fig2}.

Models of bar formation and evolution should be in agreement with these results. Conversely, models where bar properties are
imposed can use these results as a guide in the adjustment of the bar properties. One must note, however, that, due to the
relatively poor spatial resolution of the SDSS, these results are biased against bars shorter than
$L_{\rm Bar}\approx2-3$ kpc, typically
seen in very late type spirals (later than Sc \cite{elm85}). These results are thus representative of the prototypical,
bonafide bars seen mostly in early type spirals (earlier than Sc) and lenticulars. Interestingly, the median bar
ellipticity is $\approx20\%$ higher than the value found via ellipse fitting to galaxy images in \cite{mar07}.
This is exactly what was predicted in \cite{gad08} as ellipse fits systematically underestimate the true bar ellipticity
due to the dilution of the bar isophotes by the rounder, axisymmetric light distribution of bulge and disc.
When fitting ellipses to galaxy images one does not separate the contributions to the total galaxy light distribution from
the different components, but this is done in image fitting with different models for each component.
This shows that results based on the ellipticity of bars measured via ellipse fitting should be considered with
this caveat in mind. For instance, there is an indication
from ellipse fits that, for faint galaxies, disk-dominated galaxies have more
eccentric bars than bulge-dominated galaxies \cite{bar07}.
It it is not clear, however, if this result holds if the axisymmetric light contribution from bulge and disk is
taken into account. Figure \ref{fig1} also shows that most of these bars have a semi-major axis between 3 and 6 kpc
(but with a long tail to longer bars), and that bars do not extend further than $\approx3$ times the disc scalelength
$h$, or $\approx1r_{24}$ (the radius at which the galaxy surface brightness reaches
24 mag arcsec$^{-2}$ in the $r$-band --
see also \cite{erw05}). Figure \ref{fig2} shows that a typical bar is responsible for $\approx10\%$
of the total galaxy light, and has a quite flat luminosity profile. Interestingly,
one also sees that, indeed, bars have very
boxy shapes.

\begin{figure}
\centering
\includegraphics[width=5.5cm,clip=true]{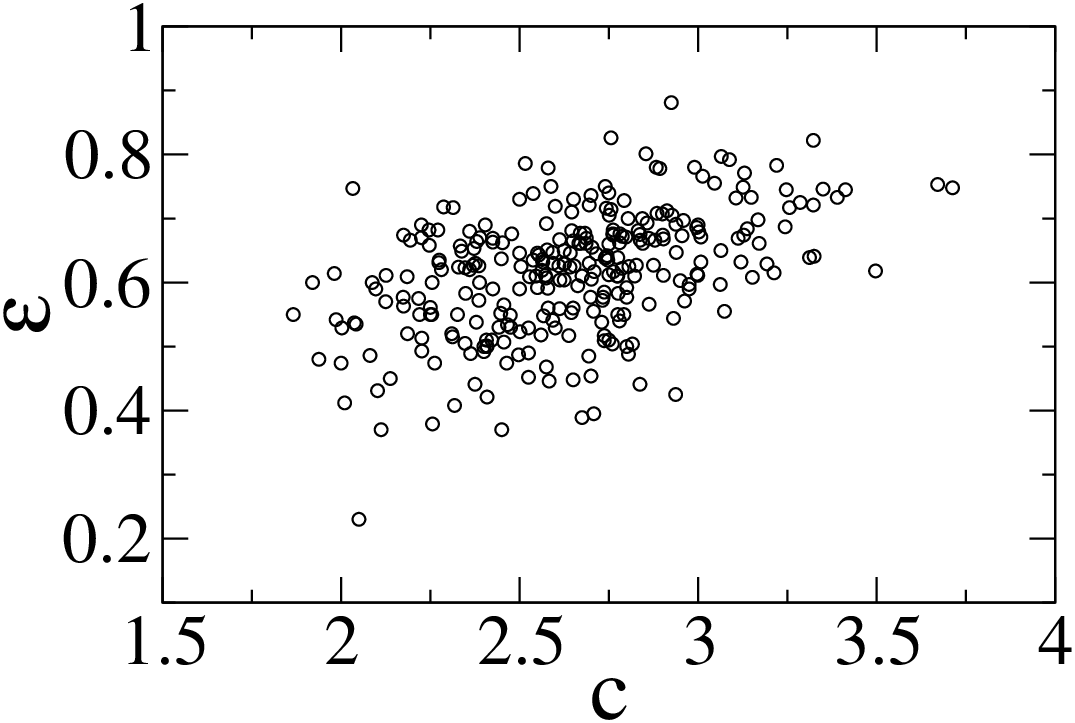}\hskip 0.2cm\includegraphics[width=5.5cm,clip=true]{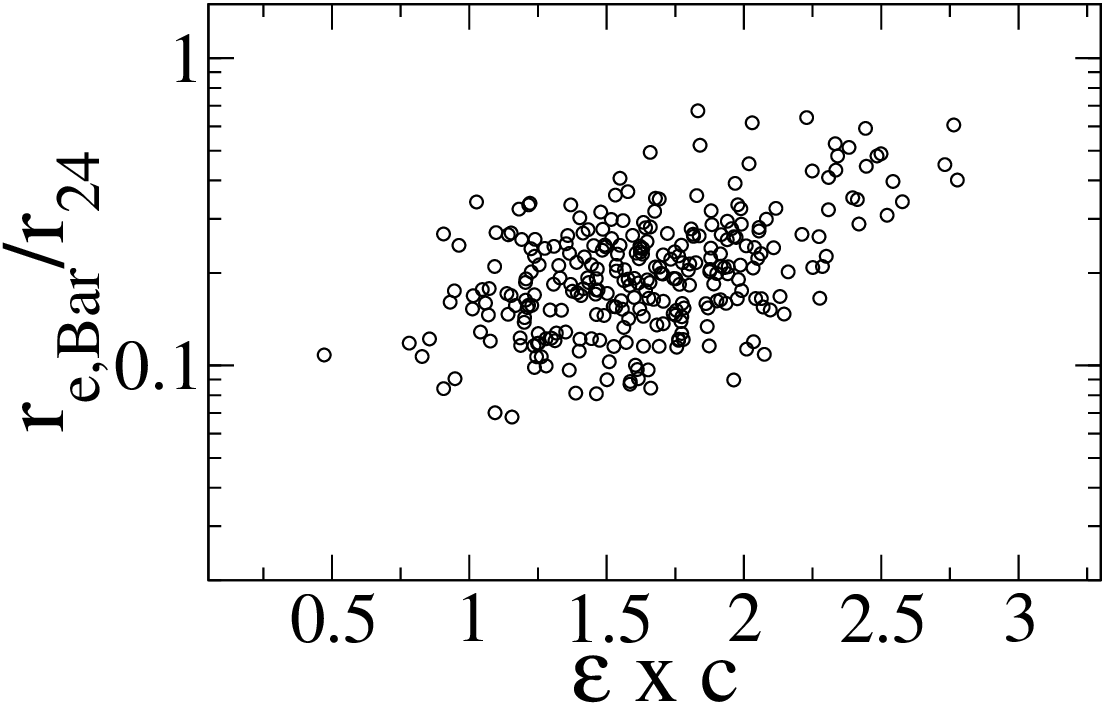}\\
\caption{Left: correlation between the bar ellipticity and boxyness. Right: correlation between the effective
radius of the bar normalised by $r_{24}$ and the product of the bar ellipticity and boxyness.}
\label{fig3}
\end{figure}

\begin{figure}
\centering
\includegraphics[height=4.1cm,clip=true]{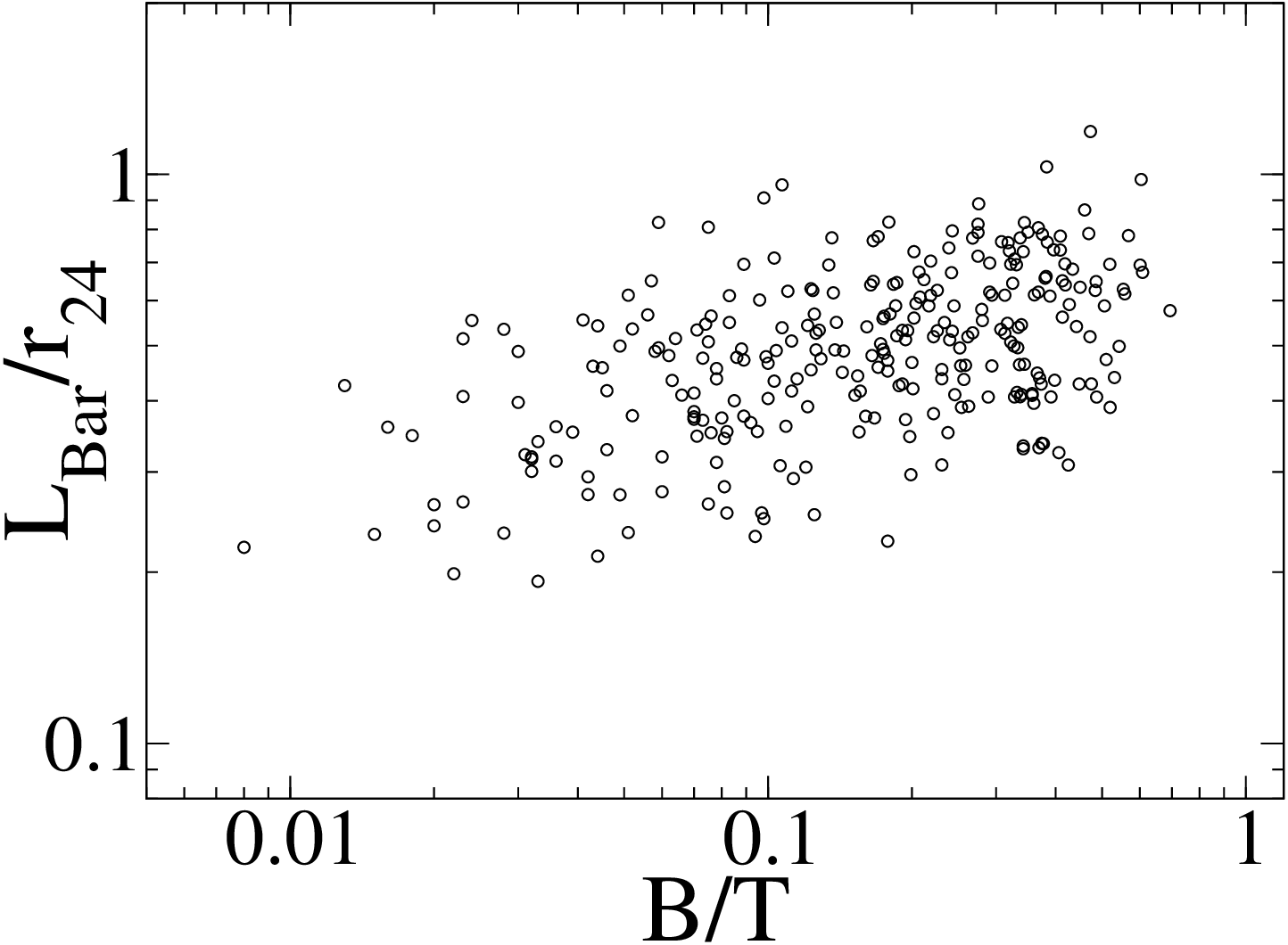}\hskip 0.2cm\includegraphics[height=4.1cm,clip=true]{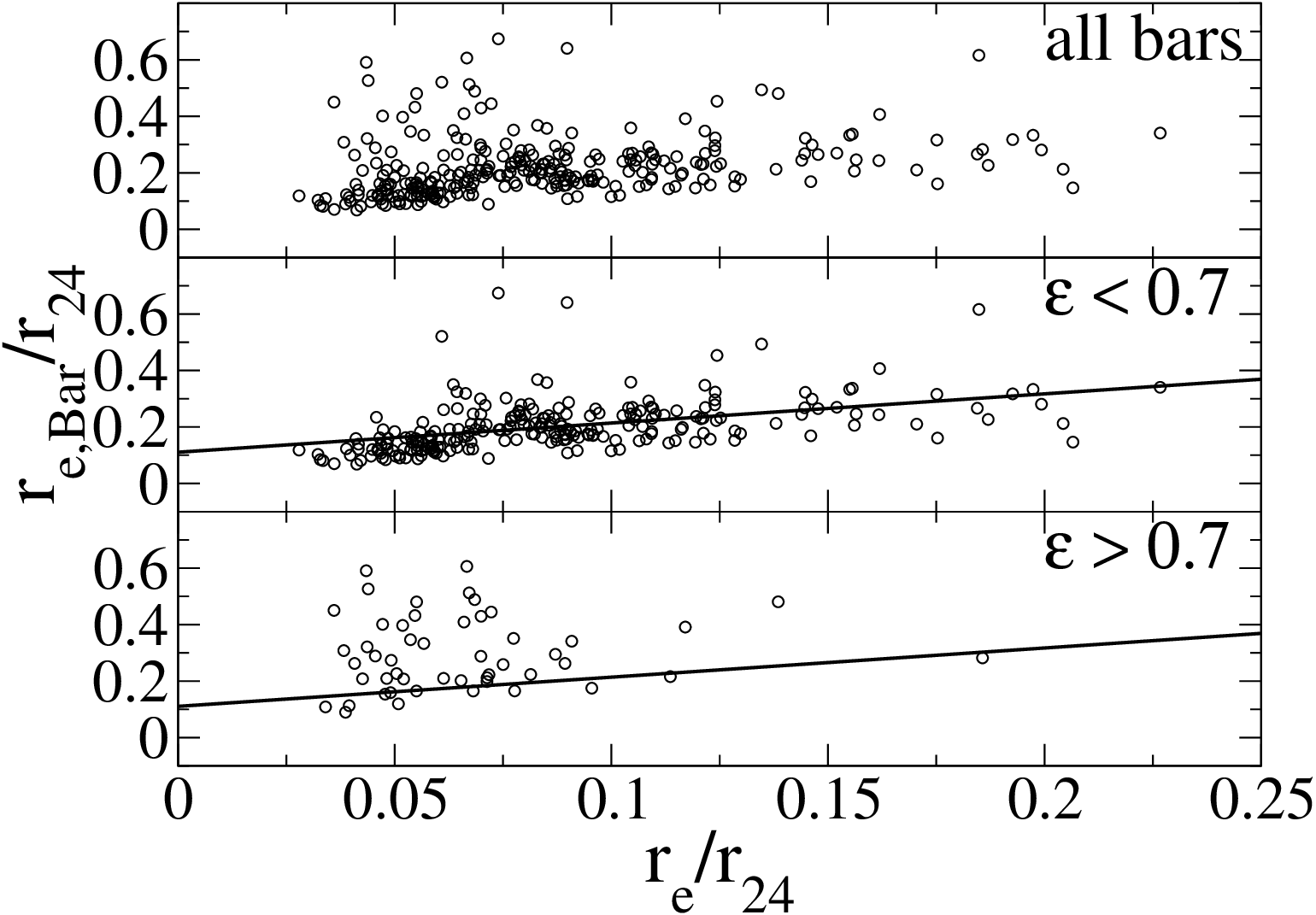}\\
\caption{Left: correlation between the length of the bar normalised by $r_{24}$ and the bulge-to-total ratio of the galaxy.
Right: correlation between the effective radius of the bar and the effective radius of the bulge, both normalised by $r_{24}$.
The top panel shows all bars, the middle panel shows bars with ellipticity $<0.7$, and the bottom panel shows bars with ellipticity
$\geq0.7$ (which, interestingly enough, include several outliers).
The solid lines are a fit to the data in the middle panel.}
\label{fig4}
\end{figure}

These data also reveal interesting correlations. The left-hand panel in Fig. \ref{fig3} shows a correlation between bar
ellipticity and boxyness, i.e. more eccentric bars are also more boxy. Both quantities contribute to the strength of the bar.
Thus, the product of both, $\epsilon\times c$, can be used as a measure of bar strength. The right-hand panel shows that the
effective radius of the bar, normalised by $r_{24}$, is correlated with $\epsilon\times c$
(see also \cite{men07}). Thus, longer bars are thinner
and stronger, as expected from the theoretical models mentioned above. The left-hand panel in Fig. \ref{fig4} shows that the
length of the bar, normalised by $r_{24}$, is correlated with the bulge-to-total ratio of the galaxy. Considering these results
and the theoretical expectations together, they indicate that bars grow longer, thinner and stronger with age, as a result
of angular momentum exchange, and that bars have had more time to evolve in galaxies with more massive bulges. In fact,
the right-hand panel in Fig. \ref{fig4} shows that the normalised effective radii of bars and bulges are correlated
(see also \cite{ath80}).
Hence, the growth of bars and bulges seems to be somehow connected. Through different paths, these conclusions are also
reached by others \cite{she07,elm07}. A more thorough analysis of these data will be published elsewhere.

\section{Future Work}
\label{sec:fut}

The connection recently found between bars and dark matter haloes opens the possibility of indirectly assessing the physical
properties of haloes through the observed properties of bars within them. With this aim, one first needs to make
a detailed comparison between real and simulated bars. If simulations can successfully reproduce the structural properties
of barred galaxies, then they might indeed give us useful estimates of physical properties of real haloes,
via comparisons of observations of barred galaxies to models with known halo properties. Such a project
has been started (see \cite{gad07}), and preliminary results are encouraging: $n$-body snapshots are being used as
real galaxy images as input to {\sc budda}, and a careful comparison of the structural parameters so obtained with those of
real barred galaxies shows that simulations are able to generally reproduce the observed quantities.

It is evident the need of further work on the methodology to estimate the dynamical ages of bars. I already mentioned above
some possible ways in this direction, and the need to enlarge the sample for which this parameter is measured. Another important
parameter that has to be measured for larger samples is the bar pattern speed
(see e.g. \cite{ger03,rau05,tre07}). Models of bar formation and evolution make
clear predictions on the behaviour of this parameter, which seems to be related to a number of other physical properties:
the central concentration of the dark matter halo, the exchange of angular momentum between disc and halo, bar age, and
even bar generation. It will be very useful to have estimates, for a large sample, of {\em both} bar age and pattern speed.

On the theory side, it is important now to obtain an updated criterion for the onset of the bar instability in discs,
accounting for the role of the halo. This will be very useful for semi-analytic models (see \cite{mo98}) which consider
bar instability based on earlier studies. Such models can also benefit largely from detailed prescriptions for
the transport of disc material to the centre, and the building of disc-like and box/peanut bulges. It is only
recently that $n$-body simulations dedicated to study the formation and evolution of bars started to use responsive,
cosmologically motivated haloes, and it is naturally expected that significant progress will come from such studies. Finally,
the observed dichotomy between bars in early and late type disc galaxies has to be better understood. It is very likely that
the availability of a large gas content in the disc plays a key role. Theoretical work focused on the effects of gas
in bar formation and evolution can substantially improve our understanding on this subject.

\begin{acknowledgement}
I am very grateful to the organisers, in particular Nikos Voglis and Panos Patsis, for this wonderful opportunity.
I benefitted from discussions with several authors, especially Lia Athanassoula and Peter Erwin. I thank
Guinevere Kauffmann for letting me present here previously
unpublished results from our structural analysis of SDSS images. Comments from the reviewer, Preben Grosb\o l, were
greatly appreciated and helpful to improve this paper.
The author is supported by the Deutsche Forschungsgemeinschaft priority program 1177 (``Witnesses of Cosmic
History: Formation and evolution of galaxies, black holes and their environment''), and the Max Planck
Society.
\end{acknowledgement}

\printindex
\end{document}